\documentclass[3p,sort&compress]{elsarticle}

\usepackage{lineno,hyperref}
\modulolinenumbers[5]
\usepackage{amssymb}
\usepackage{amsmath}
\usepackage{multirow}
\usepackage{adjustbox}
\usepackage{epsfig} 
\usepackage{threeparttable}
\usepackage{color}

\journal{Nuclear Physics A}

\def\12C{$^{12}$C}
\def\14C{$^{14}$C}
\def\15C{$^{15}$C}
\def\16C{$^{16}$C}
\def\17C{$^{17}$C}

\def\TR{$T_{\rm R}$}

\begin{document}

\begin{frontmatter}

\title{Nuclear matter distributions in the neutron-rich carbon isotopes	$^{14-17}$C from intermediate-energy proton elastic scattering in inverse
	kinematics}

\author[address1]{A.V.~Dobrovolsky\corref{correspondingauthor}}
\cortext[correspondingauthor]{Corresponding author}
\ead{Dobrovolsky\_AV@pnpi.nrcki.ru}

\author[address1]{G.A.~Korolev}
\author[address3]{S.~Tang\fnref{fn1}}
\author[address1]{G.D.~Alkhazov}
\author[address2]{G.~Col\`{o}}
\author[address3]{I.~Dillmann\fnref{fn2}}
\author[address3]{P.~Egelhof}
\author[address3]{A.~Estrad\'{e}\fnref{fn3}}
\author[address3]{F.~Farinon}
\author[address3]{H.~Geissel}
\author[address3]{S.~Ilieva}
\author[address1]{A.G.~ Inglessi}
\author[address3]{Y.~Ke\fnref{fn1}}
\author[address1]{A.V.~Khanzadeev}
\author[address3]{O.A.~Kiselev}
\author[address3]{J.~Kurcewicz\fnref{fn4}}
\author[address3]{L.X.~Chung\fnref{fn5}}
\author[address3]{Yu.A.~Litvinov}
\author[address1]{G.E.~Petrov}
\author[address3]{A.~Prochazka}
\author[address3]{C.~Scheidenberger}
\author[address1]{L.O.~Sergeev}
\author[address3]{H.~Simon}
\author[address3]{M.~Takechi\fnref{fn6}}
\author[address4]{V.~Volkov\fnref{fn7}}
\author[address1]{A.A.~Vorobyov}
\author[address3]{H.~Weick}
\author[address1]{V.I.~Yatsoura}

\address[address1]{Petersburg Nuclear Physics Institute, National Research Centre Kurchatov Institute, Gatchina, 188300 Russia}
\address[address3]{GSI Helmholtzzentrum f\"{u}r Schwerionenforschung GmbH, 64291 Darmstadt, Germany}
\address[address2]{Dipartimento di Fisica, Universit\`{a} degli Studi di Milano and INFN, Sezione di Milano, Via Celoria 16, 20133 Milano, Italy}
\address[address4]{Institut f\"{u}r Kernphysik, Technische Universit\"{a}t Darmstadt, 64289 Darmstadt, Germany}

\fntext[fn1]{Present address: Institute of Modern Physics, Chinese Academy of Sciences, 509 Nanchang Rd., Lanzhou 730000, China}
\fntext[fn2]{Present address: TRIUMF, 4004 Wesbrook Mall, Vancouver, BC V6T 2A3, Canada}
\fntext[fn3]{Present address: Department of Physics, Central Michigan University, Mount Plesant, MI 48859, USA}
\fntext[fn4]{Present address: ISOLDE, CERN, CH-1211 Geneva 23, Switzerland}
\fntext[fn5]{Present address: Department of Nuclear Physics, INST, 179 Hoang Quoc Viet, Nghia Do, Cau Giay, Ha Noi, Vietnam}
\fntext[fn6]{Present address: Graduate School of Science and Technology, Niigata University, Niigata 950-2102, Japan}
\fntext[fn7]{Present address: NRC 'Kurchatov Institute' -- ITEP, Moscow, 117218, Russia}

\begin{abstract}
	The absolute differential cross sections for small-angle proton elastic scattering off the nuclei $^{12,14-17}$C have been measured in inverse kinematics at energies near 700 MeV/u at GSI Darmstadt. 
	The hydrogen-filled ionization chamber IKAR served simultaneously as a gas target and a detector for the recoil protons. The projectile scattering angles were measured with multi-wire tracking detectors. The radial nuclear matter density distributions and the root-mean-square nuclear matter radii were deduced from the measured cross sections using the Glauber multiple-scattering theory. A possible neutron halo structure in $^{15}$C, $^{16}$C and $^{17}$C is discussed. The obtained data show evidence for a halo structure in the $^{15}$C nucleus.
\end{abstract}


\end{frontmatter}


\section{Introduction}

The study of nuclei far from stability is a topic of great current interest. A number of experiments have shown that these nuclei may have exotic structures such as a neutron skin or a halo~\cite{Tanihata96,Jonson04,Tanihata13,Ozawa01-1}. The neutron skin describes an excess of neutrons on the nuclear surface whereas the neutron halo corresponds to such an excess along with an extended tail of the neutron density distribution. The necessary conditions for the halo formation in nuclei are a small binding energy and a low angular momentum of the valence nucleon(s). It has been found that a halo structure manifests itself by large interaction (reaction) cross sections, by enhanced removal cross sections and by narrow momentum distributions of reaction products in the processes of nuclear break-up and Coulomb dissociation~\cite{Tanihata96,Jonson04,Hansen95}.

A long isotopic chain of carbon nuclei was extensively studied both experimentally and theoretically with the aim to understand the evolution of the nuclear structure as one approaches the drip line. Among other topics, the variation of the nuclear shape with the neutron excess~\cite{Ren96,Sagawa04,Kanada05}, the development of a halo~\cite{Tanihata13,Kanungo16,Bazin98,Sauvan04,Kobayashi12}, and the change of the shell structure~\cite{Tanihata13,Otsuka20,Tran18} are important subjects in the study of the nuclei of carbon isotopes. Recently, an experimental evidence for a prevalent $Z = 6$ magic number in neutron rich carbon isotopes was presented~\cite{Tran18} based on a systematic study of proton radii, electromagnetic transition rates and atomic masses of light nuclei. Small neutron separation energies are known in $^{15}$C, $^{17}$C, $^{19}$C and $^{22}$C~\cite{Wang12}, so these nuclei are suggested to be candidates to exhibit a neutron halo. Large enhancements in the values of the root-mean-square (rms) nuclear matter radius $R_{\rm m}$ evaluated from the measured interaction cross sections were found for $^{15}$C, $^{19}$C~\cite{Ozawa01-2,Kanungo16} and $^{22}$C~\cite{Togano16}. These results also signal the formation of a neutron halo. Narrow fragment momentum distributions of the reaction products in the nuclear break-up of $^{15}$C~\cite{Bazin98,Sauvan04,Fang04,Rodriges10}, $^{19}$C~\cite{Rodriges10,Baumann98,Kobayashi12} and $^{22}$C~\cite{Kobayashi12} support the existence of a halo structure in these nuclei.

The situation concerning a halo formation in $^{17}$C is rather contradictory. In several experimental studies a broad momentum distribution observed from the one-neutron nuclear break-up of $^{17}$C contradicts a halo existence in this nucleus~\cite{Bazin98,Sauvan04,Rodriges10,Baumann98}. This was explained by a \textit{d}-wave nature of the valence neutron in its ground state. The matter radius derived from the interaction cross section measured at 965 MeV/u did not show a significant enhancement in respect to its neighbours~\cite{Ozawa01-2}. On the other hand, such an enhancement was predicted in theoretical studies of the properties of the nuclear structure of carbon isotopes~\cite{Lu13} within the relativistic Hartree-Fock-Bogolubov theory. The authors of Ref.~\cite{Lu13} suggest single-neutron halo structures in both $^{17}$C and $^{19}$C nuclei. The reaction cross section for scattering of $^{17}$C on a $^{12}$C target was measured at 79 MeV/u at RIKEN~\cite{Wu04}. On the basis of the finite-range Glauber model, the density distribution in $^{17}$C was derived using the measured reaction cross section together with the interaction cross section deduced at high energy. From these results a long tail in the neutron density distribution in $^{17}$C~\cite{Wu04} was suggested. Later, the same experimental data were reanalysed~\cite{Fan14} using the well tested modified Glauber model~\cite{Takechi09}. The results of the analysis~\cite{Fan14} showed that $^{17}$C is a halo-like nucleus with a big deformation and a tail structure. The deformation may explain the broad momentum distribution of $^{16}$C fragments from $^{17}$C~\cite{Fan14}.

The information on the structure of $^{16}$C is also contradictory. In an experiment at RIKEN~\cite{Zheng02}, the reaction cross section for scattering of $^{16}$C projectiles on a $^{12}$C target was measured at an energy of 83 MeV/u. The analysis of the data suggests that $^{16}$C has a (core + 2$n$) structure and demonstrates the formation of a neutron halo~\cite{Zheng02}. This would explain an enhancement of the $^{16}$C reaction cross section at low energy~\cite{Zheng02} and an enhancement of the $^{16}$C interaction cross section measured at relativistic energy at GSI~\cite{Ozawa01-2}. The same conclusion about the halo formation in $^{16}$C was also drawn in Ref.~\cite{Rashdan12}, where a strong prolate deformation of this nucleus was predicted. However, the $^{16}$C nucleus has a relatively large neutron separation energy, $S_{\rm 2n} = 5.468$~MeV~\cite{Wang12}, which is not consistent with the existence of a halo in this nucleus. Later, it was found~\cite{Yamaguchi03} that the momentum distribution of $^{14}$C fragments from the $^{16}$C break-up is rather broad with a FWHM of $142 \pm 14$~MeV/\textit{c}, which also contradicts a halo formation in $^{16}$C. Recently new calculations on the structure of the $^{15}$C and $^{16}$C nuclei~\cite{Rashdan20} lead the author to the conclusion that $^{15}$C is a halo nucleus, while $^{16}$C has a skin structure.

Probing the nuclear matter distributions in stable nuclei with proton elastic scattering at intermediate energies near 1 GeV is known to be a well established method~\cite{Tanihata13,Alk78,Sakaguchi17}. In order to study the structure of exotic nuclei, experiments in inverse kinematics were proposed~\cite{Alk92} and performed by the PNPI-GSI collaboration at energies of secondary beams around 700 MeV/u at GSI, Darmstadt~\cite{Alk97,Neumaier02,Alk02,Egelhof02,Dobrov06,Ilieva12,Korolev18,Dobrov19}. In these experiments, the hydrogen filled ionization chamber IKAR~\cite{Alk97,Neumaier02,Vor74} was used as an active target to measure with high accuracy the absolute differential cross sections for proton elastic small-angle scattering on exotic nuclei. An analysis of the measured cross sections using the Glauber multiple scattering theory makes it possible to study the nuclear matter distributions and to determine the rms of the total matter radii and the radii of the nuclear cores and halos~\cite{Alk92,Alk02}. Previously, the method was applied to study the neutron rich nuclei $^6$He, $^8$He, $^8$Li, $^9$Li, $^{11}$Li, $^{12}$Be, $^{14}$Be~\cite{Alk97,Neumaier02,Alk02,Egelhof02,Dobrov06,Ilieva12}, and the proton rich nuclei $^7$Be and $^8$B~\cite{Korolev18,Dobrov19}. Measurements on the stable nuclei $^4$He and $^6$Li, which have equal numbers of protons ($Z$) and neutrons ($N$), and for which the difference between the neutron and proton distributions is expected to be small, were used to make a consistency check of the experimental method~\cite{Alk02,Dobrov06}.

In the present experiment, the $^{12,14,15,16,17}$C nuclei of the carbon isotopic chain were investigated employing the same method of the proton elastic scattering in inverse kinematics. The aim of the experiment was to obtain the nuclear matter density distributions in the $^{14-17}$C isotopes and to study a possible halo structure in $^{15-17}$C. The $^{14}$C nucleus was chosen as a presumable core for the $^{15}$C and $^{16}$C nuclei. The measurement of the differential cross section for elastic $p^{12}$C scattering was used as a consistency check of the experimental method, including the data analysis procedure.

\section{Experimental set-up and the measurement procedure}

\begin{figure} 
	\centering
	\psfig{file=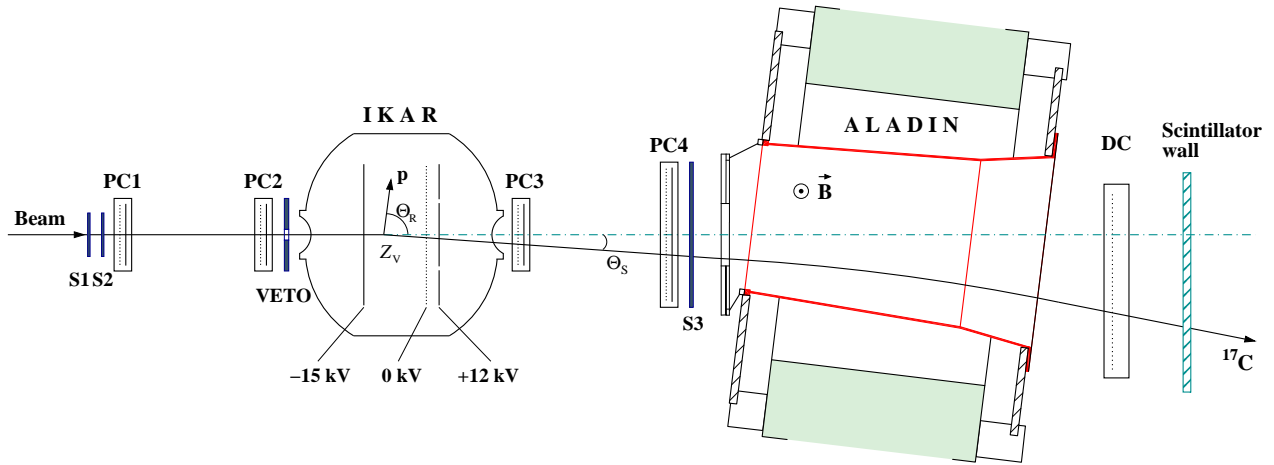,scale=0.61} 
	\caption{Schematic view of the experimental set-up. IKAR is the ionization chamber which serves as a hydrogen target and a detector of the recoil protons. Only one IKAR chamber module of six identical ones is shown. IKAR allows to determine the proton recoil energy \TR, the recoil angle $\Theta_{\rm R}$, and the vertex interaction point $Z_{\rm V}$. PC1--PC4 are multi-wire proportional chambers for measuring the projectile scattering angle $\Theta_{\rm S}$. S1--S3 and VETO are scintillator detectors for beam identification and triggering. The ALADIN magnet with the drift chamber DC and a scintillator wall are for identification of the scattered projectiles.}
	\label{setup}
\end{figure}

The measurements were performed at GSI, Darmstadt, at the exit of the fragment separator FRS~\cite{Geissel92} using the experimental set-up shown in Fig.~\ref{setup}. The carbon isotopes were produced through fragmentation of the $^{22}$Ne primary beam interacting with a 8 g/cm$^2$ thick Be target. The produced secondary beams with an energy of $\sim$700 MeV/u and an energy spread of $\sim$1.3\% were focused at the centre of the active target IKAR, the mean energies of the beam particles being determined with an accuracy of about 0.1\%. The intensity of the secondary carbon beams was at the level of 3000 ions/s with a duty cycle in the range of 50--70\%.

The experimental set-up was the same as in the previous experiment~\cite{Ilieva12}. It includes the active target IKAR~\cite{Vor74,Burq78,Vor82}, a tracking system based on multi-wire proportional chambers PC1--PC4, scintillator detectors S1--S3 and VETO, the ALADIN magnet with a drift chamber and a scintillator wall. The active target IKAR is the hydrogen-filled ionization chamber which serves as a hydrogen target and a proton recoil detector. IKAR consists of six identical cells, one of which is shown in Fig.~\ref{setup}. It permits to measure the energy $T_{\rm R}$ of the recoil proton (or its energy loss in case it leaves the active volume), the scattering angle $\Theta_{\rm R}$ of the scattered proton, and the coordinate $Z_{\rm V}$ of the interaction point along the chamber axis in the grid-cathode space~\cite{Neumaier02}.

The scattered beam particles were registered in coincidence with the recoil protons. The scattering angle $\Theta_{\rm S}$ of the projectiles was determined with a set of two-dimensional multi-wire proportional chambers PC1--PC4. The $\Theta_{\rm S}$ angular resolution was estimated to be in the range from $\sigma_{\rm{\Theta}} = 0.6$~mrad for the case of $^{17}$C to $\sigma_{\rm{\Theta}} = 0.85$~mrad for $^{12}$C.

\begin{figure}[]
	\centering
	\psfig{file=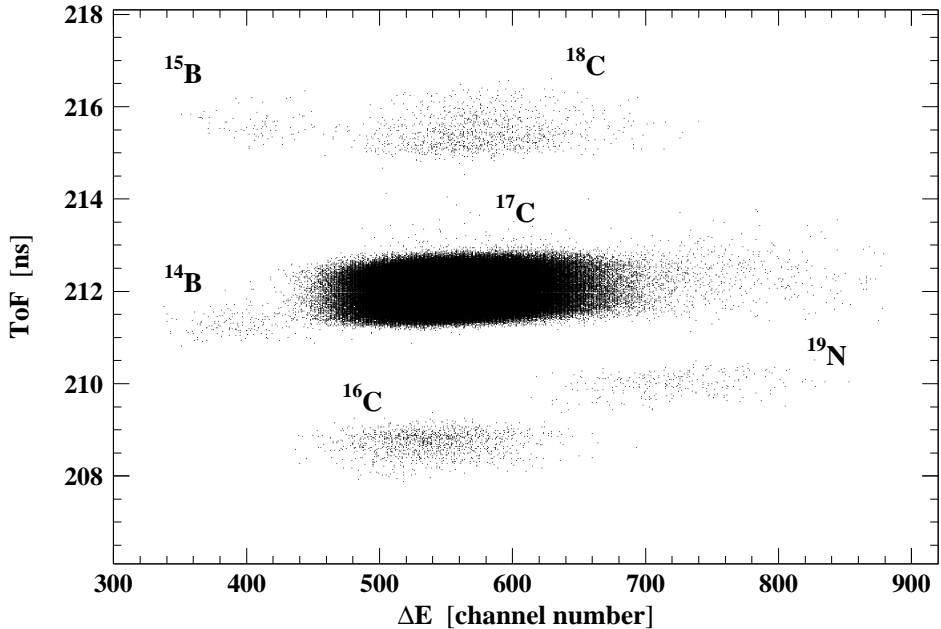,scale=0.7}
	\caption{2-Dimensional plot of the time-of-flight (ToF) and the energy loss ($\Delta E$) in the S3 scintillator for the case of the $^{17}$C beam.}
	\label{beam}
\end{figure}

A set of scintillation counters (S1, S2, S3 and VETO) was used for triggering and identification of the beam particles $via$ time-of-flight (ToF) and energy loss ($\Delta E$) measurements. The identification plot for the case of the $^{17}$C secondary beam is shown in Fig.~\ref{beam}. The time-of-flight and energy loss of the projectiles in the scintillators allow for unambiguous discrimination of the different isotopes present in the beam. The contamination with other nuclei for each selected carbon isotope was below the 0.1\% level.

The ALADIN magnet with a drift chamber and a scintillator wall behind it was utilized to discriminate against break-up reaction channels using magnetic rigidity and energy loss of the reaction products. Some features of the experimental lay-out and a detailed description of the procedure of the measurements have already been described in earlier publications~\cite{Neumaier02,Alk02,Egelhof02,Dobrov06,Ilieva12,Korolev18,Dobrov19}.

The differential cross section d$\sigma$/d$t$ was determined after the event selection using the relation
\begin{equation}
\frac{{\rm d}\sigma}{{\rm d}t} = \frac{{\rm d}N_{\rm{el}}}{{\rm d}t  N_{\rm{b}}  n  \Delta L}~.
\end{equation}
Here, d$N_{\rm{el}}$  is the number of elastic proton-nucleus scattering events in the interval ${\rm d}t$ of the four-momentum transfer squared, $N_{\rm{b}}$ is the total number of incident beam particles, $n$ is the density of protons in the target, and $\Delta L$ is the total target length. The value of $t$ was calculated as $ |t| = 2 m T_{\rm R}$, (where $m$ is the mass of the proton) for the lower momentum transfers, or as $ |t| = 4 p^2 {\rm sin}^2(\Theta_{\rm{S}}/2) / (1 + 2 E {\rm sin}^2(\Theta_{\rm{S}}/2) /mc^2)$ (where $p$ and $E$ denote the projectile initial momentum and total energy, correspondingly) for the higher momentum transfers~\cite{Dobrov19}.

The procedure of the selection of elastic events was the same as in the previous experiments with IKAR~\cite{Neumaier02,Dobrov06,Ilieva12,Dobrov19}. The measured differential cross sections are to a large extent cross sections for elastic scattering. However, they may contain some admixture of inelastic scattering. Possible contributions of inelastic scattering to the measured cross sections were estimated by calculations.

The calculations of the inelastic cross sections for proton scattering off the carbon isotopes under study were performed using the eikonal model. In particular, the formalism of Ref.~\cite{Lenzi88} was adopted as a starting point, but it was extended in order to distinguish between scattering on protons and neutrons in the nuclei under investigation. Note that for the case of neutron-rich nuclei, such a distinction is obviously necessary. In the calculations, the basic inputs were the nucleon-nucleon ($NN$) scattering amplitudes and the ground-state (transition) densities for the cases of elastic (inelastic) scattering, respectively. The parameters of the $NN$ amplitudes were taken from Ref.~\cite{Bertulani03}. The ground-state densities were described as Gaussians, while the transition densities were as in the Tassie model~\cite{Tassie56}. The rms radii of the proton and neutron distributions $R_{\rm p}$ and $R_{\rm n}$ were taken from Ref.~\cite{Kanungo16}. The total differential inelastic cross sections for the different carbon isotopes were calculated by summing up the contributions of all experimentally known states below the particle threshold~\cite{NNDC}. The deformation parameters $\beta_{\rm p}$ and $\beta_{\rm n}$ used in the calculations were based on the existing experimental information (proton or other scattering data, Coulomb excitation or electromagnetic decay properties). The details of the calculations will be published elsewhere~\cite{Colo}.

\begin{figure}[]
	\centering
	\psfig{file=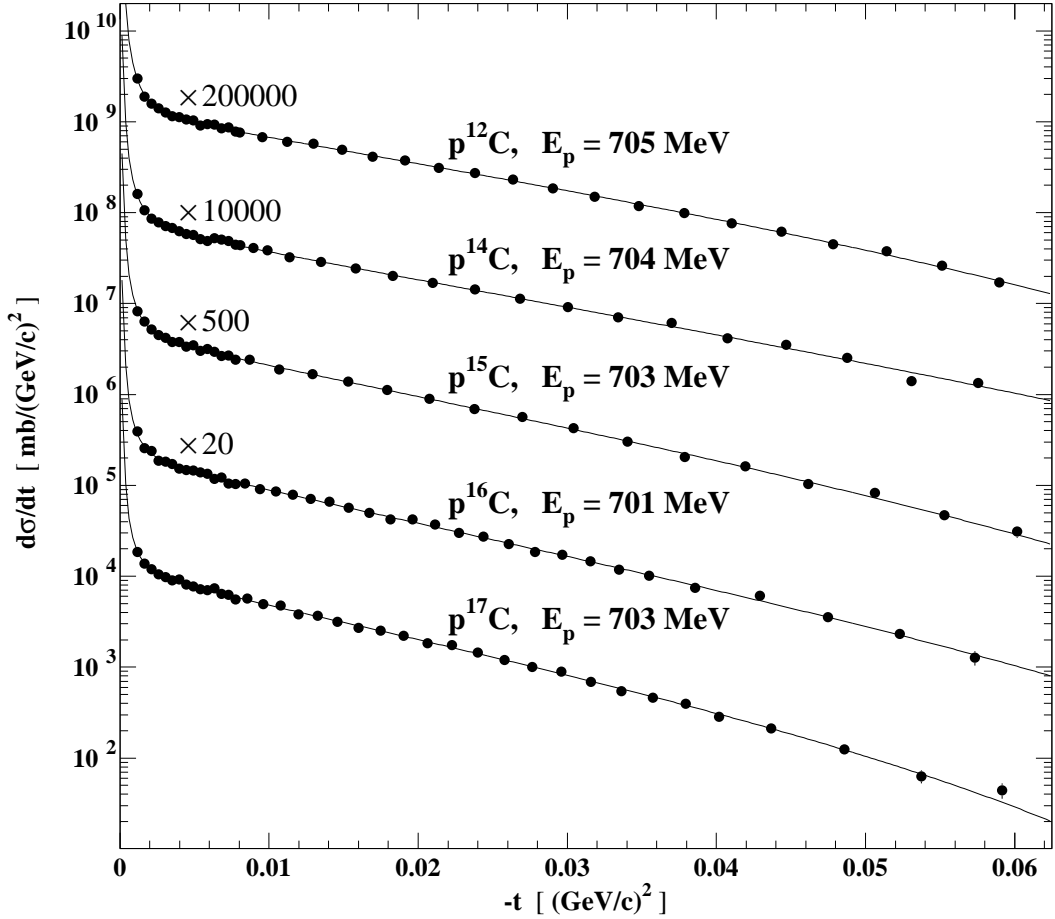,scale=0.7}
	\caption{Absolute differential cross sections d$\sigma$/d$t$ for $p^{12,14,15,16,17}$C elastic scattering versus the four-momentum transfer squared $-t$. The indicated energies correspond to the equivalent proton energies for direct kinematics. Solid lines are the results of fits to the experimental cross sections performed within the Glauber theory using the GH parameterization with the fitted parameters.} 
	\label{crs}
\end{figure}

The calculated inelastic cross sections are significantly smaller than the measured values of d$\sigma$/d$t$ and make a noticeable contribution to d$\sigma$/d$t$ (up to about 10\%) only at the highest values of $|t|$ (at $|t| \simeq 0.06$~(GeV/$c)^2$). The absolute differential cross sections d$\sigma$/d$t$ deduced in the present experiment according to Eq. (1) for proton elastic scattering on the $^{12}$C, $^{14}$C, $^{15}$C, $^{16}$C, and $^{17}$C nuclei in the momentum-transfer range of $0.002 \leq |t| \leq 0.06$~(GeV/$c)^2$ after subtraction of the calculated contributions from the inelastic scattering are displayed in Fig.~\ref{crs} and listed in a tabular form in the Appendix. The indicated energies $E_{\rm p}$ correspond to the equivalent proton energies in direct kinematics. A high detection efficiency for the beam particles and the elastic-scattering events provide the 2\% accuracy of the absolute normalization of the measured cross sections. The uncertainty in the $t$-scale calibration is estimated to be about 1.5\%. Note that the above discussed procedure of subtraction of the estimated contributions of the inelastic scattering had a rather small effect (within the error bars) on the deduced radii.

\section{The data analysis and results}

The Glauber multiple-scattering theory was used to obtain the nuclear density distributions from the measured cross sections similarly as in the previous experiments with IKAR~\cite{Alk02,Egelhof02,Dobrov06,Ilieva12,Korolev18,Dobrov19}. The calculations were performed using the basic Glauber formalism for proton-nucleus elastic scattering and taking experimental data on the elementary proton-proton and proton-neutron scattering amplitudes as input (for details see Ref.~\cite{Alk02}). In the analysis of the experimental data, the nuclear many-body density $\rho_{\rm A}$ was taken as a product of the one-body densities, which were parameterized with different functions. The parameters of these densities were found by fitting the calculated cross sections to the experimental data. The fitting procedure is described in detail in Ref.~\cite{Alk02}.

In order to reduce the model dependence of the obtained results, four parameterizations of phenomenological nuclear density distributions were applied in the present analysis, labeled as SF (Symmetrized Fermi), GH (Gaussian-Halo), GG (Gaussian-Gaussian) and GO (Gaussian-Oscillator). A detailed description of the SF, GH, GG and GO parameterizations is given in Ref.~\cite{Alk02}. Within the GH and SF density parameterizations, the many-body density is the product of the one-body densities, assuming that all nucleons have the same density distribution, while within the GG and GO parametrizations, the nuclear density is subdivided into the core and valence (\textquotedblleft halo\textquotedblright) nucleon components. The free parameters in the GG and GO parameterizations are the rms radii $R_{\rm{c}}$ and $R_{\rm{v}}$ ($R_{\rm{h}}$) of the core and valence (\textquotedblleft halo\textquotedblright) nucleon distributions. The matter radius $R_{\rm m}$ is connected with $R_{\rm{c}}$ and $R_{\rm{v}}$ by the following relation:
\begin{equation}
R_{\rm m} = [(A_{\rm c} R_{\rm c}^2 + A_{\rm v} R_{\rm v}^2) / A]^{1/2},
\end{equation}
where $A$ is the nuclear mass number, $A_{\rm c}$ is the number of nucleons in the core, and $A_{\rm v}$ is the number of valence nucleons.

\begin{table}[t]
	\centering
	\caption{Parameters obtained by fitting the calculated proton elastic scattering cross sections for the carbon isotopes under investigation to the measured ones for the parameterizations SF, GH, GG and GO of the nuclear matter density distributions. The presented parameters refer to point-nucleon density distributions. The parameters are as follows: $R_{\rm m}$ -- rms nuclear matter radius; $R_{\rm c}$ -- rms nuclear core radius; $R_{\rm v}$ -- rms radius of the valence (\textquotedblleft halo\textquotedblright) nucleon(s) distribution; $N_{\rm df}$ -- number of degrees of freedom; $R_{\rm 0}$ -- \textquotedblleft half density radius\textquotedblright\ and $a$ -- diffuseness parameter of the SF distribution; $\alpha$ -- the parameter of the GH distribution which influences the shape of the distribution (see~\cite{Alk02}); $A_{\rm n}$ -- normalization parameter of the calculated cross section. $A_{\rm n}$, $\chi ^2 / N_{\rm df}$ and $\alpha$ are dimensionless, all other fit parameters are given in fm. The radii $R_{\rm c}$ and $R_{\rm v}$ are in the c.m. system of the nucleus. All errors given are statistical only.}
	\label{fit_results}
	\vspace*{2pt}
	\begin{tabular}{cccllll}
		\hline
		\rule[1ex]{0pt}{1.0ex}	
		\multirow{2}{*}{Nucleus}          & \multirow{2}{*}{Parameterization} & \multicolumn{1}{c}{\multirow{2}{*}{$\chi^2/N_{\rm df}$}} &                         \multicolumn{3}{c}{Fit parameters}& \multicolumn{1}{c}{\multirow{2}{*}
			{\begin{tabular}[c]{@{}c@{}}					$R_{\rm m}$, \\
					fm
			 \end{tabular}}} \\ \cline{4-6}
		&        &      \multicolumn{1}{c}{}   &  \multicolumn{1}{c}{\rule[1ex]{0pt}{1.5ex} $A_{\rm n}$}  &  \multicolumn{2}{c}{\rule[1ex]{0pt}{1.5ex}	Density parameters}  &  \\
		\hline
		\multicolumn{1}{c}{	\rule[1ex]{0pt}{1.5ex}	\multirow{2}{*}{$^{12}$C}} &                SF                &                     30.0/33                     & 1.03(1)                         & $R_{\rm 0}$ = 1.98(13) & a = 0.48(3)            & 2.35(2)                             \\
		\multicolumn{1}{l}{}            &                GH                &                     30.2/33                     & 1.03(1)                         & $R_{\rm m}$ = 2.33(1)   & $\alpha$ = 0.00(2)     & 2.33(1)                             \\
		\multicolumn{1}{l}{}            &       \multicolumn{1}{l}{}       &                                                 &                                 &                         &                        &  \\
		\multirow{2}{*}{$^{14}$C}            &                SF                &                     31.1/31                     & 1.01(1)                         & $R_{\rm 0}$ = 0.87(32)  & a = 0.63(3)            & 2.43(2)                             \\
		&                GH                &                     31.4/31                     & 1.01(1)                         & $R_{\rm m}$ = 2.41(2)   & $\alpha$ = 0.11(2)     & 2.41(2)                             \\
		\multicolumn{1}{l}{}            &       \multicolumn{1}{l}{}       &                                                 &                                 &                         &                        &  \\
		\multirow{4}{*}{$^{15}$C}            &                SF                &                     32.6/29                     & 1.03(1)                         & $R_{\rm 0}$ = 1.56(16)  & a = 0.62(2)            &
		2.59(2)                             \\
		&                GH                &                     32.6/29                     & 1.03(1)                         & $R_{\rm m}$ = 2.57(2)   & $\alpha$ = 0.06(2)     & 2.57(2)                            
		\\
		&                GG                &                     34.4/29                     & 1.02(1)                         & $R_{\rm c}$ = 2.43(1)   & $R_{\rm v}$ = 4.45(43) & 2.61(5)                            
		\\
		&                GO                &                     33.6/29                     & 1.02(1)                         & $R_{\rm c}$ = 2.40(1)   & $R_{\rm v}$ = 4.49(33) & 2.60(4)                             
		\\
		\multicolumn{1}{l}{}            &       \multicolumn{1}{l}{}       &                                                 &                                 &                         &                        &  \\
		\multirow{4}{*}{$^{16}$C}            &                SF                &                     33.5/37                     & 1.04(1)                         & $R_{\rm 0}$ = 1.31(25)  & a = 0.67(3)            &
		2.70(3)                             \\
		&                GH                &                     36.3/37                     & 1.04(1)                         & $R_{\rm m}$ = 2.68(3)   & $\alpha$ = 0.09(2)     & 2.68(3)                            
		\\
		&                GG                &                     35.3/37                     & 1.04(1)                         & $R_{\rm c}$ = 2.43(2)   & $R_{\rm v}$ = 4.36(29) & 2.75(6)                            
		\\
		&                GO                &                     35.0/37                     & 1.04(1)                         & $R_{\rm c}$ = 2.38(2)   & $R_{\rm v}$ = 4.35(22) & 2.71(4)                             
		\\
		\multicolumn{1}{l}{}            &       \multicolumn{1}{l}{}       &                                                 &                                 &                         &                        &  \\
		\multirow{4}{*}{$^{17}$C}            &                SF                &                     35.0/37                     & 1.01(1)                         & $R_{\rm 0}$ = 1.97(13)  & a = 0.60(2)            &
		2.69(2)                             \\
		&                GH                &                     34.7/37                     & 1.02(1)                         & $R_{\rm m}$ = 2.67(2)   & $\alpha$ = 0.03(2)     & 2.67(2)                            
		\\
		&                GG                &                     35.5/37                     & 1.02(1)                         & $R_{\rm c}$ = 2.58(2)   & $R_{\rm v}$ = 3.86(54) & 2.68(3)                            
		\\
		&                GO                &                     35.3/37                     & 1.02(1)                         & $R_{\rm c}$ = 2.56(2)   & $R_{\rm v}$ = 4.06(40) & 2.67(3)                             
		\\		\hline
	\end{tabular}
\end{table}

The results of the fits to the measured experimental cross sections with the phenomenological density distributions SF, GH, GG and GO for the carbon isotopes under investigation are presented in Table~\ref{fit_results}. For each density parameterization, the deduced rms nuclear matter radius $R_{\rm m}$, the $\chi^2$ value of the fitting procedure, the values of the fit parameters, and the normalization coefficient $A_{\rm n}$ with which the calculated cross section d$\sigma$/d$t$ was multiplied to obtain the same absolute normalization as the experimental one are presented.
Note that the errors in Table~\ref{fit_results} are statistical only.

The solid lines in Fig.~\ref{crs} represent the results for the cross sections d$\sigma$/d$t$ calculated using the GH parameterization with the fitted parameters. At $|t| < 0.005$~(GeV/$c)^2$, a steep rise of the cross section with decreasing $|t|$ is caused by Coulomb scattering. It is seen that the fits describe the experimental cross sections fairly well with the reduced $\chi^2$ values close to 1.0. The calculations of the cross sections with the nuclear matter density parameterizations SF, GG, and GO with the fitted parameters give  practically the same results.

For the description of the cross sections in the case of the $^{12}$C and $^{14}$C nuclei, only the SF and GH density parameterizations were used. The weighted mean values of $R_{\rm m}$ averaged over the results obtained with these density parameterizations are:
\begin{center} 
	$R_{\rm m} = (2.34 \pm 0.05)$ fm  \hspace{1.5cm}for $^{12}$C,
	
	$R_{\rm m} = (2.42 \pm 0.05)$ fm  \hspace{1.5cm}for $^{14}$C.
\end{center}

\noindent The errors indicated here and below for the deduced values of the radii include statistical and systematic uncertainties~\cite{Alk02}. The systematic errors appear as the result of uncertainties in the absolute normalization of the experimental cross sections, as an error in the $t$-scale and errors introduced to the analysis from uncertainties in the parameters of the free $pp$ and $pn$ scattering amplitudes. Also, the contributions to the systematic errors due to corrections for the inelastic scattering and due to different model density parameterizations used are taken into account.

In the analysis it was assumed that the nuclei $^{15}$C and $^{17}$C consist of the $^{14}$C and $^{16}$C cores, respectively, and a loosely bound valence neutron. For these nuclei good descriptions of the cross sections have been achieved with all the density parameterizations used. The corresponding values of the rms matter radii $R_{\rm m}$ deduced with all four parameterizations for $^{15}$C and $^{17}$C are close to each other within rather small errors. The values of $R_{\rm m}$ averaged over the results obtained with all the density parameterizations are:
\begin{center} 
	$R_{\rm m} = (2.59 \pm 0.05)$ fm  \hspace{1.5cm}for $^{15}$C,
	
	$R_{\rm m} = (2.68 \pm 0.05)$ fm  \hspace{1.5cm}for $^{17}$C.
\end{center}

The mean value for the core radius of $^{15}$C deduced with the GG and GO parameterizations is $R_{\rm c} = 2.41 (5)$~fm. Combining the obtained values of $R_{\rm m}$ and $R_{\rm c}$, and employing relation (2) between the rms radii $R_{\rm m}$, $R_{\rm c}$, and $R_{\rm v}$, one derives for the radius of the valence neutron distribution in $^{15}$C a value of $R_{\rm v} = 4.36 (38)$ fm. The mean values of the core radius and the radius of the valence neutron distribution deduced in the present analysis for $^{17}$C are $R_{\rm c} = 2.57 (5)$ fm and $R_{\rm v} = 4.05 (47)$ fm.

In the analysis of the data for the $^{16}$C nucleus with the density parameterization within the GG and GO models a structure of a $^{14}$C core plus two valence neutrons was assumed. For this isotope all density parameterizations also fit the experimental data well. The weighted mean rms matter radius of $^{16}$C, deduced from the GH, SF, GG, and GO parameterizations is

\begin{center} 
	$R_{\rm m} = (2.70 \pm 0.06)$ fm.
\end{center}

\noindent For the core radius and the radius of the valence neutrons distribution, the following mean values were determined: $R_{\rm c} = 2.41 (5)$ fm and $R_{\rm v} = 4.20 (26)$ fm.

\begin{figure}[]
	\centering
	\psfig{file=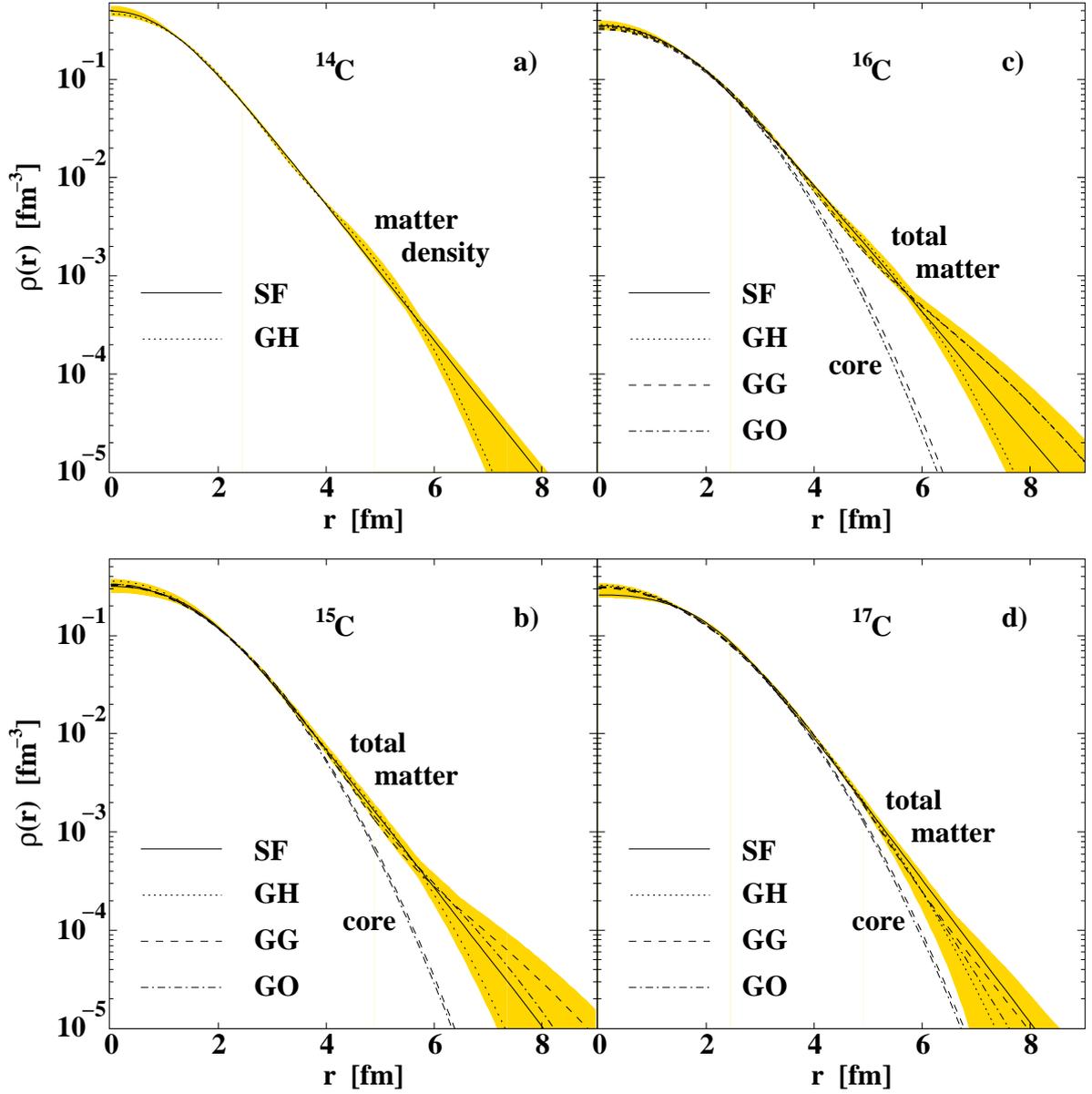,scale=0.85}
	\caption{Total and core matter distributions $\rho (r)$ of the nuclear density in $^{14}$C (a), $^{15}$C (b), $^{16}$C (c) and $^{17}$C (d) deduced in the analysis by using model density parameterizations SF (Symmetrized Fermi), GH (Gaussian-Halo), GG (Gaussian-Gaussian), and GO (Gaussian-Oscillator), for details see the text. The shaded areas represent the envelopes of the density variation within the model parameterizations applied, superimposed by the statistical errors. All density distributions are normalized to the number of nucleons.} 
\label{dens}
\end{figure}

The deduced nuclear matter density distributions obtained using different parameterizations of the nuclear matter distributions are plotted in Fig.~\ref{dens}. The shaded areas represent the envelopes of the density variation within the model parameterizations applied, superimposed by the statistical errors. Figure~\ref{dens} also shows the obtained core matter distributions. All density distributions refer to point-nucleon distributions.

Using the matter radii $R_{\rm m}$ deduced in the present work and the radii $R_{\rm p}$ of proton distributions obtained in Refs.~\cite{Angeli13} and \cite{Kanungo16}, the radii $R_{\rm n}$ of neutron distributions and thicknesses of the neutron skins $\delta_{\rm np} = R_{\rm n} - R_{\rm p}$ for the nuclei of the studied carbon isotopes were determined (see Table~\ref{rad_comp}) with the help of expression (3):
\begin{equation}
R_{\rm n} = [(A R_{\rm m}^2 - Z R_{\rm p}^2) / N]^{1/2}.
\end{equation}

\section{Discussion}

Recently, the charge-changing cross sections for the $^{12-19}$C nuclei were measured at GSI at 900 MeV/u with a carbon target by Kanungo $et$ $al$.~\cite{Kanungo16}. Using a finite-range Glauber model, the authors derived radii $R_{\rm p}$ of the proton density distributions for the studied carbon isotopes. With these values of $R_{\rm p}$ fixed, they performed a new analysis of the interaction cross sections from Ref.~\cite{Ozawa01-2} to obtain more accurate values of the matter radii $R_{\rm m}$. The authors also performed coupled-cluster computations using chiral nucleon-nucleon and three-nucleon interactions which satisfactorily describe the experimental data on proton and matter radii.

\begin{figure}[h]
	\centering
	\psfig{file=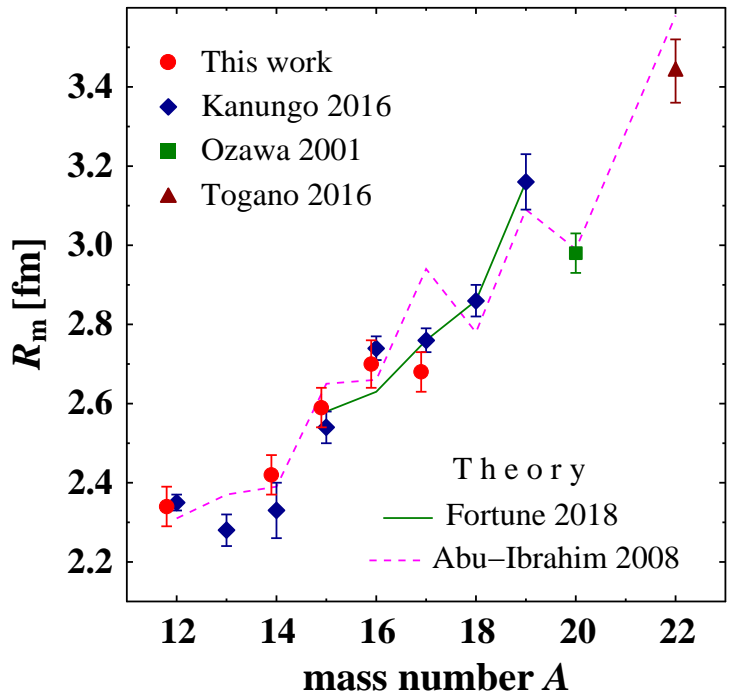,scale=0.75}
	\caption{Nuclear matter radii of carbon isotopes. Experimental data are: this work (circles), the results of \cite{Kanungo16} (diamonds), the result of \cite{Ozawa01-1} (square), and the result of \cite{Togano16} (triangle). Theoretical predictions are taken from \cite{Fortune18} (solid line) and \cite{Abu08} (dashed line).}
	\label{radii}
\end{figure}

Our results on $R_{\rm m}$ for the carbon isotopes are compared with the results of Ref.~\cite{Kanungo16} in Table~\ref{rad_comp} and in Fig.~\ref{radii}. It is seen that the present results on $R_{\rm m}$ turn out to be within the experimental errors in agreement with the results of Ref.~\cite{Kanungo16}. In Fig.~\ref{radii} are also shown experimental results of Refs.~\cite{Ozawa01-1,Togano16} and two sets of theoretical predictions for the matter radii of the carbon isotopes~\cite{Fortune18,Abu08}. The matter radii in \cite{Fortune18,Abu08} were calculated using a simple model under the assumption that the considered nuclei consist of a core plus one or two valence neutrons. Note that the radii calculated in \cite{Abu08} exhibit a pronounced staggering effect $-$ the radii for the odd mass numbers are larger than the average of the radii for the neighbour even mass numbers.

\begin{table}[b]
	\centering
	\caption{Comparison of the present results on the rms radii of the nuclear matter with the values derived from the Glauber analysis of interaction cross sections \cite{Kanungo16}. In addition data on the radii of the proton distribution $R_{\rm p}$ (from Refs. \cite{Kanungo16} and \cite{Angeli13}) and on the deduced radii of the neutron distribution $R_{\rm n}$ and on the thickness of the neutron skin $\delta_{\rm np}$ are presented.}
	\label{rad_comp}
	\vspace*{2pt}
	\begin{tabular}{cccccll}
		\hline 
		\rule[1ex]{0pt}{1.5ex}	\multirow{2}{*}{Isotope} & $R_{\rm m}$, fm & $R_{\rm m}$, fm & \multicolumn{2}{c}{$R_{\rm p}$, fm} & $R_{\rm n}$, fm & $\delta_{\rm np}$, fm \\ 
		\rule[-1ex]{0pt}{2.5ex}  & This work & Ref. \cite{Kanungo16} & Ref. \cite{Angeli13} & Ref. \cite{Kanungo16} & This work & This work \\ 
		\hline 
		\rule[1ex]{0pt}{2.0ex} $^{12}$C & 2.34 (5) & 2.35 (2) & 2.34 (1) &  & 2.34 (10) & 0.00 (10) \\ 
		\rule[-1ex]{0pt}{2.5ex} $^{14}$C & 2.42 (5) & 2.33 (7) & 2.38 (2) &  & 2.45 (9) & 0.07 (9) \\ 
		\rule[-1ex]{0pt}{2.5ex} $^{15}$C & 2.59 (5) & 2.54 (4) &  & 2.37 (3) & 2.73 (8) & 0.36 (9) \\ 
		\rule[-1ex]{0pt}{2.5ex} $^{16}$C & 2.70 (6) & 2.74 (3) &  & 2.40 (4) & 2.86 (9) & 0.46 (10) \\ 
		\rule[-1ex]{0pt}{2.5ex} $^{17}$C & 2.68 (5) & 2.76 (3) &  & 2.42 (4) & 2.81 (8) & 0.39 (9) \\ 
		\hline 
	\end{tabular} 
\end{table}

The method applied in the given work to study the nuclear matter density distributions was previously tested with the data on proton scattering from stable nuclei $^4$He~\cite{Alk02} and $^6$Li~\cite{Dobrov06}. The differential cross section for $p^{12}$C elastic scattering measured in this work was also used to check the method. The $^{12}$C matter radius $R_{\rm m} = 2.34 (5)$ fm derived in the present work is in agreement with the value of $R_{\rm m} = 2.35 (2)$ fm of Ref.~\cite{Kanungo16}. Note that the rms charge radius of $^{12}$C is known with high precision~\cite{Angeli13} from $e^-$ scattering and muonic $x$-ray measurements: $R_{\rm ch} = 2.470 (2)$ fm.
Taking into account the finite size effect of the nucleon (see, $e.g.$, Ref.~\cite{Tanihata13}) and the value of the proton charge radius $r_{\rm p} = 0.8414 (19)$ fm \cite{Nist}, the rms radius $R_{\rm p}$ of the proton distribution in $^{12}$C is obtained to be $R_{\rm p} = 2.34 (1)$ fm. The number of neutrons in $^{12}$C is equal to that of protons, therefore the matter and proton distributions (normalized to one nucleon) are expected to be rather similar. Indeed, the $R_{\rm m}$ value deduced in the present work has occurred to be equal to the value of $R_{\rm p}$ extracted from the experimental data on the charge radius of $^{12}$C. This result on $p^{12}$C scattering demonstrates a consistency check of the present experimental method, including the procedure of the data analysis.

The $^{14}$C nucleus is of interest as the presumable core in $^{15}$C and $^{16}$C \cite{Tran18}. This nucleus is supposed to have a spherical shape due to the neutron closed shell effect~\cite{Ren96,Sagawa04,Kanada05}. The present value of $R_{\rm m} = 2.42 (5)$ fm is in agreement within errors with the result $R_{\rm m} = 2.33 (7)$ fm of Ref.~\cite{Kanungo16}. The charge radius $R_{\rm ch} = 2.503 (9)$ fm \cite{Angeli13} of $^{14}$C may be used to find the corresponding radius of the proton distribution $R_{\rm p} = 2.38 (2)$ fm. By combining the matter radius $R_{\rm m}$, deduced in the present work for $^{14}$C with the value of $R_{\rm p}$, and using expression (3), the rms radius of the neutron distribution $R_{\rm n}$ in $^{14}$C has been determined to be $R_{\rm n} = (2.45 \pm 0.09)$ fm. Thus, within the error bars, the $^{14}$C nucleus has the same radius of the neutron distribution $R_{\rm n}$ as that of the proton distribution $R_{\rm p}$: $R_{\rm n} \approx R_{\rm p}$.

The structure of the odd isotope $^{15}$C has been considered in a ($^{14}$C-core + $n$) model. This nucleus has a small neutron separation energy $S_{\rm n} = 1.218$ MeV, so it is suggested to be a candidate for a halo nucleus. A special feature of the present method is that it makes possible to determine the sizes of the nuclear core and of the halo. The ratio of the determined valence nucleon to the core nucleon radius, $\kappa = R_{\rm v} /R_{\rm c}$, may be used as a gauge for the halo existence \cite{Grigorenko98}. Theory predicts typically values of $\kappa \leq 1.25$ for light nuclei near the valley of beta stability, while for a halo structure this value can be $\kappa \approx 2$, or even larger \cite{Jonson04}. In the present analysis, a value of $\kappa = 1.81$ for $^{15}$C is obtained, which confirms the suggestion \cite{Tanihata13} that this nucleus demonstrates a \textquotedblleft moderate halo formation\textquotedblright.

Due to the low binding energy of the halo neutron in $^{15}$C, it is natural to expect that the internal core size $R_{\rm c}^*$ (size of the core in its own c.m. system) is close to that of the free $^{14}$C nucleus. The motion of the c.m. of the core around the c.m. of the whole nucleus slightly increases the effective core size 
$R_{\rm c}$ \cite{Alk02}. Following Tanihata $et$ $al$. \cite{Tanihata13}, the internal core size $R_{\rm c}^*$ in the (core + $n$) model turns out to be
\begin{equation}
R_{\rm c}^* = (R_{\rm c}^2 - \rho_{\rm c}^2)^{1/2},
\end{equation}
\noindent where $\rho_{\rm c}$ is the rms distance between the c.m. of the core and the c.m. of the whole nucleus:
\begin{equation}
\rho_{\rm c} = R_{\rm v} / (A -1).
\end{equation}

In the present analysis we obtain $\rho_{\rm c} = 0.31 (3)$ fm and $R_{\rm c}^* = 2.39 (5)$ fm for $^{15}$C. The latter value agrees with $R_{\rm m} = 2.42 (5)$ fm for $^{14}$C. Taking for $^{15}$C the proton radius $R_{\rm p} = (2.37 \pm 0.03)$ fm~\cite{Kanungo16}, and using Eq. (3), the rms neutron radius for $^{15}$C is determined to be $R_{\rm n} = (2.73 \pm 0.08)$ fm, and for the thickness of the neutron skin we deduce the value of $\delta_{\rm np} = (0.36 \pm 0.09)$ fm (see Table~\ref{rad_comp}).

There are several theoretical considerations of the structure of $^{16}$C, which is treated as a ($^{14}$C-core $+n+n$) three-body system~\cite{Fortune18,Abu08}. The experimental value of $R_{\rm m} = 2.70 (6)$ fm, deduced in the present work for $^{16}$C, is in good agreement with existing experimental data as well as with theoretical results (Fig.~\ref{radii} and Table~\ref{rad_comp}). The core size in $^{16}$C ($R_{\rm c} = 2.41 (5)$ fm) is close to the size of the free $^{14}$C nucleus ($R_{\rm m} = 2.42 (5)$ fm). According to the present analysis, the ratio of the valence nucleon radius $R_{\rm v}$ to the core radius $R_{\rm c}$ turns out to be equal in $^{16}$C to $\kappa = 1.74$, which is smaller than the $\kappa$ values of the $2n$ halo nuclei $^{11}$Li ($\kappa = 2.71$~\cite{Dobrov06}) and $^{14}$Be ($\kappa = 1.91$~\cite{Ilieva12}) determined earlier with the same method. This observation suggests that the spatial distribution of two valence neutrons in $^{16}$C should be considered rather as a skin, than as a halo.
Using the matter radius of the present work $R_{\rm m} = (2.70 \pm 0.06)$ fm and the radius of the proton distribution $R_{\rm p} = (2.40 \pm 0.04)$ fm \cite{Kanungo16}, we obtain for the radius of the neutron distribution $R_{\rm n} = (2.86 \pm 0.09)$ fm, and for the thickness of the neutron skin, the value $\delta_{\rm np} = (0.46 \pm 0.10)$ fm has been deduced (see Table \ref{rad_comp}). This result is an indication of a noticeable neutron skin in $^{16}$C.

We have considered the spatial structure of the $^{17}$C nucleus in a ($^{16}$C-core + $n$) model. The neutron separation energy $S_{\rm n}$ for $^{17}$C is small: 
$S_{\rm n} = 0.728$ MeV. Therefore, one could expect $^{17}$C to be a halo nucleus. However, the ratio of the valence nucleon radius to the core radius, determined in the present work for $^{17}$C, occurs to be relatively small, $\kappa = 1.58$, which does not support the picture that $^{17}$C is a halo nucleus. With the determined value of $R_{\rm v}$ and Eqs. (4) and (5) in the case of $^{17}$C we obtain $\rho_{\rm c} = 0.25 (3)$ fm and $R_{\rm c}^* = 2.56 (5)$ fm. This value of $R_{\rm c}^*$ is smaller than $R_{\rm m} = 2.70 (6)$ fm for the free $^{16}$C nucleus. This result demonstrates a noticeable contraction of the $^{16}$C cluster inside $^{17}$C. Obviously, $^{17}$C is a more dense nucleus than $^{16}$C. It was already supposed in Ref. \cite{Tanihata13} that the configuration of the nucleus $^{17}$C is more complicated than that in the (core + $n$) model.

\section{Summary}

The proton-nucleus elastic scattering at intermediate energies is an efficient method for the investigation of nuclear matter density distributions. In the present work, we have applied this method in inverse kinematics for the investigation of the nuclear radial structure of carbon isotopes. The absolute differential cross sections d$\sigma$/d$t$ were measured as a function of the four-momentum transfer squared $-t$ in the range $0.001 \leq |t| \leq 0.06$ (GeV/$c)^2$ for proton elastic scattering on the $^{12,14,15,16,17}$C nuclei. The cross sections were determined using secondary beams with energies near 700 MeV/u produced with the fragment separator FRS at GSI. The active target IKAR was used as a recoil-proton detector. The scattered projectiles were registered with a system of multi-wire proportional chambers, scintillation detectors, and a magnetic analysis. The analysis of the experimental data was performed using the Glauber multiple-scattering theory. The nuclear matter radii and the radial nuclear matter distributions for the carbon isotopes were determined from the measured cross sections d$\sigma$/d$t$. A good description of the experimental cross section is obtained with four phenomenological parameterizations of the nuclear density distributions (SF, GH, GG, and GO). Each of these parameterizations has two free parameters. Our results on the matter radii $R_{\rm m}$ for the studied carbon isotopes are in agreement within the experimental errors with those of Ref.~\cite{Kanungo16} evaluated from the measured interaction and charge-changing cross sections. The density distribution parameters ($R_{\rm m}$, $R_{\rm p}$) for $^{12}$C are well established values from measurements of the interaction cross sections and the charge radii. Therefore, the results on $p^{12}$C scattering were used as a consistency check of the present experimental method, including the procedure of the data analysis.

The measured cross sections are described fairly well within the (core + $n$) model for $^{15}$C and $^{17}$C, and the (core + $2n$) model for $^{16}$C. It was shown that the size of the $^{14}$C-core in the $^{15}$C and $^{16}$C nuclei is close to that of the free $^{14}$C nucleus.

A quantitative description of the halo structure for $^{15,16,17}$C was performed in the analysis of the nuclear matter distributions in these nuclei. The ratio of the valence nucleon to the core nucleon radius $\kappa = R_{\rm v} /R_{\rm c}$ was used as a gauge for the halo existence, where a value of $\kappa \gtrsim 2$ is expected for a halo nucleus.

The present analysis describes $^{15}$C as a halo nucleus with $\kappa = 1.82$, while $^{16}$C ($\kappa = 1.74$) and $^{17}$C ($\kappa = 1.58$) are considered as nuclei with a noticeable neutron skin. This conclusion is in agreement with the investigation of fragmentation reactions using radioactive carbon beams. Note that a~ narrow fragment momentum distribution as a signature of an extended valence nucleon density distribution in a halo nucleus was observed in the considered here carbon isotopes only for $^{15}$C~\cite{Bazin98,Sauvan04,Fang04,Rodriges10}, whereas broad fragment momentum distributions for $^{16}$C \cite{Yamaguchi03} and $^{17}$C \cite{Bazin98,Sauvan04,Rodriges10,Baumann98} imply no halo formation in these nuclei.

Besides the determination of the nucleon density distributions and their parameters, the precise data obtained for the differential proton elastic-scattering cross sections allow a sensitive test of theoretical predictions on the structure of the neutron-rich carbon nuclei. For this purpose, the nuclear density distributions obtained from various theoretical approaches may be used as an input to the Glauber multiple-scattering theory. Then the calculated elastic-scattering cross sections should be compared to the experimental data as it was done in Refs.~\cite{Alk02,Egelhof02,Dobrov06}.

\section*{Acknowledgements}

The authors would like to thank A.~Bleile, G.~Ickert, A.~Br\"{u}nle, K.-H.~Behr and W.~Niebur for the technical assistance and their help in the preparation and running of the experiment. The visiting group from PNPI thanks the GSI authorities for the hospitality.



\clearpage

\section*{Appendix}
The measured cross sections d$\sigma$/d$t$ for $p^{12,14-17}$C elastic scattering as a function of the four-momentum transfer squared $-t$. The indicated errors are statistical only.

\vspace{10pt}
\centering
\begin{tabular}{ccccc}
	\hline
	\multicolumn{2}{c}{\rule[1ex]{0pt}{1.5ex}	$p^{12}$C, $E_{\rm p}$=705.2 MeV}  &  &  \multicolumn{2}{c}{\rule[1ex]{0pt}{1.5ex}	$p^{12}$C, $E_{\rm p}$=705.2 MeV}  \\ 
	\hline
	\rule[1ex]{0pt}{1.5ex}	$-t$, (GeV/$c)^2$ & d$\sigma$/d$t$,  mb/(GeV/$c)^2$ &  & $-t$, (GeV/$c)^2$ & d$\sigma$/d$t$,  mb/(GeV/$c)^2$ \\ 
	\hline
	\rule[1ex]{0pt}{1.5ex}	
    0.00117      &         14965. $\pm$ 297.3        &  &      0.01300      &         2858.1 $\pm$ 67.1         \\
    0.00164      &         9489.3 $\pm$ 229.4        &  &      0.01490      &         2448.0 $\pm$ 60.5         \\
    0.00211      &         7942.5 $\pm$ 208.5        &  &      0.01694      &         2061.6 $\pm$ 54.3         \\
    0.00258      &         7060.1 $\pm$ 195.8        &  &      0.01910      &         1871.3 $\pm$ 50.7         \\
    0.00305      &         6335.2 $\pm$ 185.1        &  &      0.02140      &         1549.8 $\pm$ 45.3         \\
    0.00352      &         5742.1 $\pm$ 175.8        &  &      0.02382      &         1358.2 $\pm$ 41.7         \\
    0.00399      &         5620.2 $\pm$ 173.9        &  &      0.02636      &         1160.9 $\pm$ 38.0         \\
    0.00446      &         5290.9 $\pm$ 168.8        &  &      0.02904      &          924.7 $\pm$ 33.5         \\
    0.00493      &         5171.5 $\pm$ 166.9        &  &      0.03185      &          745.1 $\pm$ 29.8         \\
    0.00540      &         4517.4 $\pm$ 156.2        &  &      0.03478      &          589.2 $\pm$ 26.3         \\
    0.00586      &         4713.5 $\pm$ 159.9        &  &      0.03785      &          495.9 $\pm$ 24.0         \\
    0.00633      &         4636.6 $\pm$ 160.4        &  &      0.04104      &          383.7 $\pm$ 21.0         \\
    0.00680      &         4250.1 $\pm$ 155.3        &  &      0.04437      &          309.6 $\pm$ 18.9         \\
    0.00727      &         4317.7 $\pm$ 155.3        &  &      0.04782      &          223.6 $\pm$ 16.1         \\
    0.00774      &         3883.2 $\pm$ 147.3        &  &      0.05141      &          188.6 $\pm$ 14.9         \\
    0.00804      &         3793.6 $\pm$ 84.1         &  &      0.05513      &          131.3 $\pm$ 12.6         \\
    0.00956      &         3400.9 $\pm$ 77.7         &  &      0.05897      &           85.3 $\pm$ 10.4         \\
    0.01122      &         3023.1 $\pm$ 71.0         &  &                   &									\\
    \hline
\end{tabular}

\vspace{12pt}

\begin{tabular}{ccccc}
	\hline
	\multicolumn{2}{c}{\rule[1ex]{0pt}{1.5ex}	$p^{14}$C, $E_{\rm p}$ = 704.4 MeV}  &  & \multicolumn{2}{c}{\rule[1ex]{0pt}{1.5ex}	$p^{14}$C, $E_{\rm p}$ = 704.4 MeV}  \\ 
	\hline
	\rule[1ex]{0pt}{1.5ex}	$-t$, (GeV/$c)^2$ & d$\sigma$/d$t$,  mb/(GeV/$c)^2$ &  & $-t$, (GeV/$c)^2$ & d$\sigma$/d$t$,  mb/(GeV/$c)^2$ \\ 
	\hline
    \rule[1ex]{0pt}{1.5ex}
    0.00117      &        16137. $\pm$ 435.8          &  &      0.00989      &        3859.5 $\pm$ 126.6        \\
    0.00164      &        10641. $\pm$ 322.6          &  &      0.01137      &        3242.1 $\pm$  78.9        \\
    0.00211      &        8626.2 $\pm$ 284.4          &  &      0.01350      &        2855.9 $\pm$  71.7        \\
    0.00258      &        7779.0 $\pm$ 254.7          &  &      0.01581      &        2434.2 $\pm$  64.4        \\
    0.00305      &        7125.5 $\pm$ 245.8          &  &      0.01830      &        2024.8 $\pm$  57.9        \\
    0.00352      &        6813.3 $\pm$ 241.4          &  &      0.02096      &        1686.9 $\pm$  50.9        \\
    0.00399      &        6227.4 $\pm$ 228.9          &  &      0.02381      &        1434.8 $\pm$  46.7        \\
    0.00446      &        5802.7 $\pm$ 223.0          &  &      0.02683      &        1131.7 $\pm$  39.6        \\
    0.00493      &        5678.4 $\pm$ 218.2          &  &      0.03004      &         908.8 $\pm$  35.3        \\
    0.00540      &        5115.2 $\pm$ 209.8          &  &      0.03342      &         705.7 $\pm$  30.4        \\
    0.00586      &        4870.5 $\pm$ 211.9          &  &      0.03699      &         614.9 $\pm$  28.6        \\
    0.00633      &        5206.6 $\pm$ 224.0          &  &      0.04074      &         413.7 $\pm$  23.8        \\
    0.00680      &        5071.9 $\pm$ 218.3          &  &      0.04467      &         352.4 $\pm$  23.0        \\
    0.00727      &        4894.1 $\pm$ 203.3          &  &      0.04878      &         253.5 $\pm$  18.2        \\
    0.00774      &        4424.1 $\pm$ 180.1          &  &      0.05307      &         140.5 $\pm$  14.6        \\
    0.00807      &        4368.8 $\pm$ 139.0          &  &      0.05755      &         133.9 $\pm$  12.8        \\
    0.00896      &        4081.9 $\pm$ 132.7          &  &                   &									\\
    \hline
\end{tabular}
\newpage
\begin{tabular}{ccccc}
	\hline
	\multicolumn{2}{c}{\rule[1ex]{0pt}{1.5ex}	$p^{15}$C, $E_{\rm p}$ = 702.5 MeV}  &  & \multicolumn{2}{c}{\rule[1ex]{0pt}{1.5ex}	$p^{15}$C, $E_{\rm p}$ = 702.5 MeV}  \\ 
	\hline
	\rule[1ex]{0pt}{1.5ex}	$-t$, (GeV/$c)^2$ & d$\sigma$/d$t$,  mb/(GeV/$c)^2$ &  & $-t$, (GeV/$c)^2$ & d$\sigma$/d$t$,  mb/(GeV/$c)^2$ \\ 
	\hline
	\rule[1ex]{0pt}{1.5ex}
    0.00117      &       16475.9 $\pm$ 362.3          &  &      0.01069      &        3769.1 $\pm$  89.2        \\
    0.00164      &       12658.6 $\pm$ 322.1          &  &      0.01290      &        3332.1 $\pm$  80.9        \\
    0.00211      &       10386.9 $\pm$ 291.0          &  &      0.01532      &        2758.2 $\pm$  71.2        \\
    0.00258      &        8961.4 $\pm$ 269.9          &  &      0.01793      &        2253.4 $\pm$  62.6        \\
    0.00305      &        8421.4 $\pm$ 261.6          &  &      0.02075      &        1780.6 $\pm$  54.2        \\
    0.00352      &        7541.3 $\pm$ 247.6          &  &      0.02377      &        1379.4 $\pm$  46.7        \\
    0.00399      &        7553.8 $\pm$ 248.2          &  &      0.02699      &        1125.2 $\pm$  41.3        \\
    0.00446      &        6746.6 $\pm$ 235.1          &  &      0.03042      &         850.1 $\pm$  35.3        \\
    0.00493      &        6971.2 $\pm$ 239.6          &  &      0.03405      &         607.0 $\pm$  29.4        \\
    0.00540      &        6003.9 $\pm$ 222.9          &  &      0.03789      &         410.5 $\pm$  23.9        \\
    0.00586      &        6323.3 $\pm$ 229.7          &  &      0.04193      &         325.4 $\pm$  21.0        \\
    0.00633      &        5916.3 $\pm$ 221.1          &  &      0.04617      &         206.0 $\pm$  16.6        \\
    0.00680      &        5276.6 $\pm$ 208.6          &  &      0.05063      &         165.0 $\pm$  14.7        \\
    0.00727      &        5385.0 $\pm$ 215.6          &  &      0.05529      &          94.3 $\pm$  11.2        \\
    0.00774      &        4831.4 $\pm$ 206.5          &  &      0.06016      &          62.0 $\pm$   9.1        \\
    0.00869      &        4818.7 $\pm$ 104.4          &  &                   &									\\
    \hline
\end{tabular}

\vspace{12pt}

\begin{tabular}{ccccc}
	\hline
	\multicolumn{2}{c}{\rule[1ex]{0pt}{1.5ex}	$p^{16}$C, $E_{\rm p}$ = 700.5 MeV}  &  & \multicolumn{2}{c}{\rule[1ex]{0pt}{1.5ex}	$p^{16}$C, $E_{\rm p}$ = 700.5 MeV}  \\ 
	\hline
	\rule[1ex]{0pt}{1.5ex}	$-t$, (GeV/$c)^2$ & d$\sigma$/d$t$,  mb/(GeV/$c)^2$ &  & $-t$, (GeV/$c)^2$ & d$\sigma$/d$t$,  mb/(GeV/$c)^2$ \\ 
	\hline
	\rule[1ex]{0pt}{1.5ex}
    0.00117      &       19706.1 $\pm$ 495.4          &  &      0.01405      &        3295.4 $\pm$ 136.0        \\
    0.00164      &       12894.0 $\pm$ 412.0          &  &      0.01535      &        2828.7 $\pm$ 123.8        \\
    0.00211      &       11955.1 $\pm$ 394.6          &  &      0.01671      &        2488.2 $\pm$ 114.3        \\
    0.00258      &        9308.8 $\pm$ 346.6          &  &      0.01813      &        2115.6 $\pm$ 103.8        \\
    0.00305      &        9144.8 $\pm$ 343.3          &  &      0.01961      &        2099.7 $\pm$ 102.3        \\
    0.00352      &        8549.8 $\pm$ 331.8          &  &      0.02114      &        1863.2 $\pm$ 95.0         \\
    0.00399      &        7662.0 $\pm$ 314.4          &  &      0.02273      &        1496.5 $\pm$ 84.1         \\
    0.00446      &        7337.3 $\pm$ 307.9          &  &      0.02438      &        1368.8 $\pm$ 79.5         \\
    0.00493      &        7317.3 $\pm$ 308.0          &  &      0.02609      &        1133.0 $\pm$ 71.5         \\
    0.00540      &        6907.9 $\pm$ 300.3          &  &      0.02785      &         922.9 $\pm$ 63.9         \\
    0.00586      &        6707.5 $\pm$ 296.5          &  &      0.02967      &         856.7 $\pm$ 61.1         \\
    0.00633      &        5881.3 $\pm$ 276.2          &  &      0.03155      &         729.5 $\pm$ 55.8         \\
    0.00680      &        6069.1 $\pm$ 277.1          &  &      0.03349      &         592.8 $\pm$ 49.9         \\
    0.00727      &        5263.5 $\pm$ 265.4          &  &      0.03548      &         507.3 $\pm$ 45.9         \\
    0.00774      &        5162.7 $\pm$ 266.9          &  &      0.03858      &         374.0 $\pm$ 27.7         \\
    0.00838      &        5235.1 $\pm$ 188.7          &  &      0.04291      &         304.7 $\pm$ 24.7         \\
    0.00940      &        4564.0 $\pm$ 173.9          &  &      0.04748      &         178.3 $\pm$ 18.9         \\
    0.01047      &        4302.8 $\pm$ 165.1          &  &      0.05228      &         115.4 $\pm$ 15.1         \\
    0.01161      &        3954.3 $\pm$ 154.7          &  &      0.05732      &          63.6 $\pm$ 11.5         \\
    0.01280      &        3543.2 $\pm$ 143.6          &  &                   &									\\
    \hline
\end{tabular}

\begin{tabular}{ccccc}
	\hline
	\multicolumn{2}{c}{\rule[1ex]{0pt}{1.5ex}	$p^{17}$C, $E_{\rm p}$ = 703.2 MeV}  &  & \multicolumn{2}{c}{\rule[1ex]{0pt}{1.5ex}	$p^{17}$C, $E_{\rm p}$ = 703.2 MeV}  \\
	\hline
	\rule[1ex]{0pt}{1.5ex}	$-t$, (GeV/$c)^2$ & d$\sigma$/d$t$,  mb/(GeV/$c)^2$ &  & $-t$, (GeV/$c)^2$ & d$\sigma$/d$t$,  mb/(GeV/$c)^2$ \\
	\hline
	\rule[1ex]{0pt}{1.5ex}
    0.00117      &       18437.1 $\pm$ 429.9          &  &      0.01460      &        3160.4 $\pm$ 108.3         \\
    0.00164      &       13783.9 $\pm$ 361.8          &  &      0.01601      &        2718.3 $\pm$  98.8         \\
    0.00211      &       12008.3 $\pm$ 335.5          &  &      0.01748      &        2508.4 $\pm$  93.5         \\
    0.00258      &       10474.2 $\pm$ 312.1          &  &      0.01901      &        2210.0 $\pm$  86.5         \\
    0.00305      &        9801.4 $\pm$ 301.2          &  &      0.02062      &        1830.3 $\pm$  77.8         \\
    0.00352      &        9018.8 $\pm$ 288.7          &  &      0.02228      &        1758.0 $\pm$  75.4         \\
    0.00399      &        9179.1 $\pm$ 291.5          &  &      0.02401      &        1443.9 $\pm$  67.6         \\
    0.00446      &        8061.6 $\pm$ 273.2          &  &      0.02581      &        1200.3 $\pm$  61.1         \\
    0.00493      &        7765.3 $\pm$ 268.8          &  &      0.02766      &        1007.4 $\pm$  55.6         \\
    0.00540      &        7172.2 $\pm$ 258.9          &  &      0.02959      &         892.8 $\pm$  51.9         \\
    0.00586      &        7054.6 $\pm$ 257.8          &  &      0.03158      &         689.5 $\pm$  45.5         \\
    0.00633      &        7343.3 $\pm$ 266.4          &  &      0.03363      &         543.5 $\pm$  40.2         \\
    0.00680      &        6387.0 $\pm$ 252.1          &  &      0.03575      &         461.0 $\pm$  37.0         \\
    0.00727      &        6230.8 $\pm$ 245.6          &  &      0.03794      &         394.7 $\pm$  34.1         \\
    0.00774      &        5526.5 $\pm$ 231.5          &  &      0.04019      &         285.1 $\pm$  29.1         \\
    0.00853      &        5705.9 $\pm$ 160.4          &  &      0.04369      &         210.8 $\pm$  17.7         \\
    0.00961      &        4955.8 $\pm$ 147.2          &  &      0.04858      &         125.0 $\pm$  13.8         \\
    0.01076      &        4738.9 $\pm$ 140.7          &  &      0.05374      &          62.9 $\pm$  10.1         \\
    0.01198      &        3829.4 $\pm$ 123.7          &  &      0.05916      &          44.4 $\pm$   8.5         \\
    0.01326      &        3681.4 $\pm$ 119.0          &  &                   &									 \\
    \hline
\end{tabular}

\clearpage

\end{document}